\documentclass[12pt]{iopart}
\usepackage{feynmf}
\usepackage{epsfig}
\usepackage{hyperref}

\def\bea{\begin{equation}}
\def\eea{\end{equation}}
\def\bea{\begin{eqnarray}}
\def\eea{\end{eqnarray}}
\def\ba{\begin{array}}
\def\ea{\end{array}} 
\def\slash#1{\setbox0=\hbox{$#1$}#1\hskip-\wd0\dimen0=5pt\advance
       \dimen0 by-\ht0\advance\dimen0 by\dp0\lower0.5\dimen0\hbox
         to\wd0{\hss\sl/\/\hss}}

\begin{document}

\begin{flushright}
WITS-MITP-008
\end{flushright}

\title[Beyond the SM]{Some theories beyond the Standard Model}

\author{Alan S. Cornell}

\address{National Institute for Theoretical Physics, School of Physics and Mandelstam Institute for Theoretical Physics,
University of the Witwatersrand, Johannesburg, Wits 2050, South Africa}
\ead{alan.cornell@wits.ac.za}
\vspace{10pt}
\begin{indented}
\item[]February 2015
\end{indented}

\begin{abstract}
A brief review on the physics beyond the Standard Model is given, as was presented in the High Energy Particle Physics workshop on the $12^{th}$ of February 2015 at the iThemba North Labs. Particular emphasis is given to the Minimal Supersymmetric Standard Model, with mention of extra-dimensional theories also.
\end{abstract}

\section{Introduction}

\par The Standard Model (SM) of elementary particle physics, developed in the 1970's, describes the behaviour of all known elementary particles with impressive accuracy, and it all traces back to the relatively simple mathematical laws within the formalism of quantum field theory. As such, symmetry groups play a particularly important role in this context, where our three generations of quarks and leptons have their interactions mediated by the gauge bosons. However, the most mysterious field within the SM is the Higgs field, which forms a condensate filling the whole Universe. The motion of all particles is influenced by this condensate, which is how the particles gain their mass. Recall that the Higgs boson, a spin-0 particle produced from excitations of the Higgs field, was recently discovered, completing the particle spectrum of the SM, but that doesn't mean the story is now over. The study of the Higgs boson's couplings is now extremely important, and is connected to beyond the SM (BSM) physics.

\begin{figure}[h]
\centering
\includegraphics[width=5cm]{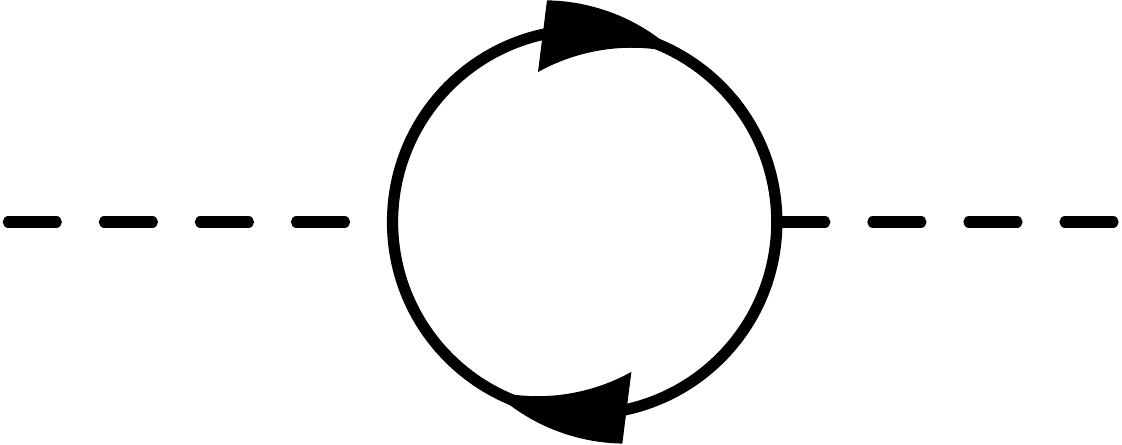} \hspace{1cm}
\includegraphics[width=4cm]{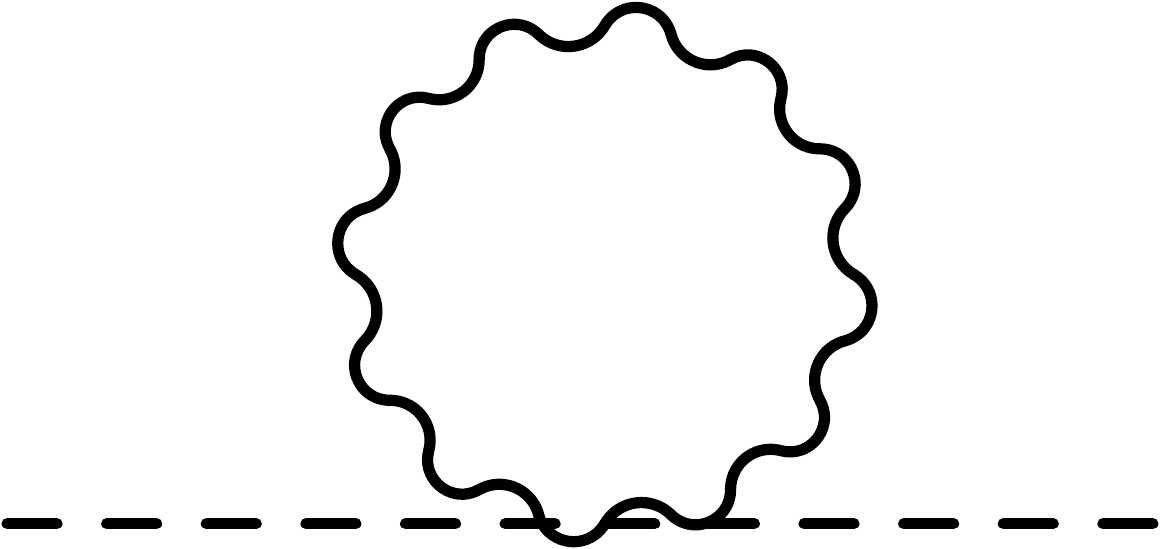} \hspace{1cm}
\includegraphics[width=4cm]{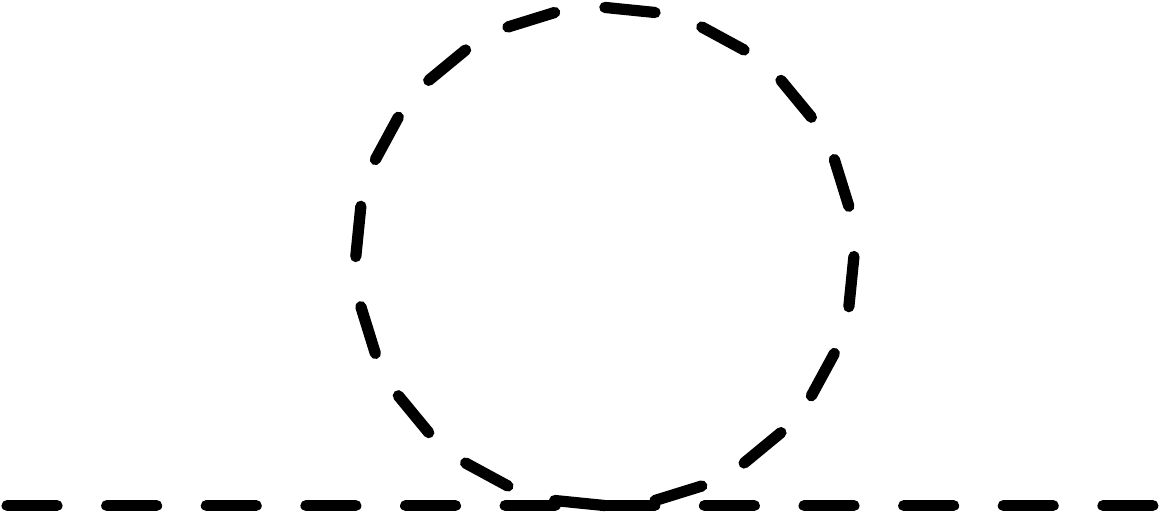}
\caption{\it The one-loop contributions to the Higgs mass, where the left diagram contains the fermion loops, the middle contributions are from the gauge bosons, and the right from Higgs self-interactions.}\label{fig:1}
\end{figure}

\par The structure of the radiative corrections to the Higgs boson is quite different from those of fermions and gauge bosons. To one-loop the renormalised Higgs mass is:
\begin{eqnarray}
m^2_h &=& m^2_{h0} - \frac{3}{8 \pi^2} y_t^2 \Lambda^2 + \frac{1}{16 \pi^2} g^2 \Lambda^2 + \frac{1}{16 \pi^2} \lambda^2 \Lambda^2 \; , \label{eqn:1}
\end{eqnarray}
where $\Lambda$ is the scale at which the loop integrals are cut off. As such, the corrections to the Higgs mass, which are referred to as ``quadratic divergences", are related to $\Lambda$. If $\Lambda$ is high, the correction is extremely large with respect to the on-shell Higgs mass. This is the ``fine-tuning problem", where such quadratic divergences are not radiative corrections to fermion or gauge boson masses (these being protected by chiral and gauge symmetries).

\par If there is an intermediate scale at which new physics manifests, this problem is resolved and the radiative corrections from any new particles ameliorates issues arising from the SM. Note that merely adding new particles that couple to the Higgs boson, to cancel one loop level quadratic corrections, is insufficient as such chance cancellations do not guarantee cancellations to all orders in a perturbative theory. To guarantee cancellations of these quadratic divergences we need a new symmetry. Some of the established approaches to handle this, which shall be reviewed here are:
\begin{itemize}
\item[1.] {\bf Supersymmetry} (SUSY): a symmetry between bosons and fermions, where the chiral symmetry of fermions controls the Higgs mass divergences by cancelling the diagrams containing SM fields with diagrams featuring their sparticles (their supersymmetric partners).
\item[2.] {\bf Extra-dimensions}: whilst we seem to reside in a four-dimensional spacetime, we may actually reside in five or more spactime dimensions. The additional or extra-dimensions are somehow ``inaccessible" to us (perhaps being compactified in some way), meaning that the true Planck mass could be of comparable scale to the electroweak scale.
\end{itemize}

\par Current experimental measurements greatly constrain each of these models, where the SM remains our most successful model to date. However, we may find deviations in the future if we focus on the few fundamental parameters present in the SM, such as the gauge couplings and the vacuum expectation value (vev) of the Higgs; as the measurements of these parameters becomes more precise, greater constraints can be placed on any BSM theory.

\par Returning to the quadratic divergences above, the divergence in Eq. (\ref{eqn:1}) is estimated by $\Lambda$, a cut-off we introduced to the loop integrals. As this momentum cut-off is independent of the external momenta, the divergences are regularisation dependent objects. For dimensional regularisation these divergences are then trivially zero. As such is the fine-tuning problem something which should be taken seriously?

\newpage
\section{Supersymmetry}

\par This fine-tuning problem, as presented above with a momentum cut-off, is valid when at some higher scale a theory with a larger symmetry exists. For the case of SUSY models our regularisation scheme must satisfy our SUSY and we cannot remove all quadratic divergences. Therefore, the radiative corrections to the Higgs sector are proportional to the mass scale of the sparticles (the SUSY scale) under a proper regularisation. Taking the sparticle masses as being much greater than the SM particle masses, the theory appears as the SM at low energies with $\Lambda$ being at the SUSY scale. This is an example where fine-tuning arguments hold for theories with an extended scale above which a new symmetry arises.

\par SUSY is a symmetry which exchanges fermions for bosons, and vice-versa. The Minimal Supersymmetric SM (MSSM) is where we extend the SM by having a SUSY in the limit when particle masses are negligible, and where this is presumed to be an effective theory of some fully supersymmetric model. As the full theory undergoes a spontaneous SUSY breaking, the sparticles obtain a mass that is much greater than those for their partner SM particles. The fermion's superpartner is a spin-0 particle (a sfermion), whilst a gauge boson's superpartner has spin-1/2 (a gaugino). For the Higgs boson, its superpartner has spin-1/2 also (a higgsino). In Tab.~\ref{tab:1} the particle content of the MSSM is listed, where the superpartners have the same charges as their SM counter-parts. This is due to the generators of the SUSY transform commuting with the $SU(3)\times SU(2)\times U(1)$ transformation of the SM. Note also that in the MSSM we have two Higgs doublets and two higgsinos, as chiral fermions with charge $(1,2)_{\pm 1/2}$ are constrained by anomaly cancellation.

\begin{table}
\caption{\label{tab:1}The additional particle content of the MSSM.}
\begin{indented}
\item[]\begin{tabular}{@{}|c|c|c|c|c|} 
\hline
Names & Spin & $P_R$ & Gauge Eigenstates & Mass Eigenstates \\
\hline
Higgs bosons & 0 & +1 & $H_u^0 \; H_d^0 \; H_u^+ \; H_d^-$ & $h^0 \; H^0 \; A^0 \; H^\pm$\\
\hline
& & & $\tilde{u}_L \; \tilde{u}_R \; \tilde{d}_L \; \tilde{d}_R$ & (same) \\
squarks & 0 & -1 &  $\tilde{s}_L \; \tilde{s}_R \; \tilde{c}_L \; \tilde{c}_R$ & (same) \\
& & & $\tilde{t}_L \; \tilde{t}_R \; \tilde{b}_L \; \tilde{b}_R$ & $\tilde{t}_1 \; \tilde{t}_2 \; \tilde{b}_1 \; \tilde{b}_2$\\
\hline
& & & $\tilde{e}_L \; \tilde{e}_R \; \tilde{\nu}_e$ & (same) \\
sleptons & 0 & -1 &  $\tilde{\mu}_L \; \tilde{\mu}_R \; \tilde{\nu}_\mu$ & (same) \\
& & & $\tilde{\tau}_L \; \tilde{\tau}_R \; \tilde{\nu}_\tau$ & $\tilde{\tau}_1 \; \tilde{\tau}_2 \; \tilde{\nu}_\tau$\\
\hline
neutralinos & 1/2 & -1 & $\tilde{B}^0 \; \tilde{W}^0 \; \tilde{H}_u^0 \; \tilde{H}_d^0$ & $\tilde{N}_1 \; \tilde{N}_2 \; \tilde{N}_3 \; \tilde{N}_4$ \\
\hline
charginos & 1/2 & -1 & $\tilde{W}^\pm \; \tilde{H}_u^+ \; \tilde{H}_d^-$ & $\tilde{C}_1^\pm \; \tilde{C}_2^\pm$ \\
\hline
gluino & 1/2 & -1 & $\tilde{g}$ & (same) \\
\hline
goldstino & 1/2 & -1 & & (same) \\
(gravitino) & 3/2 & -1 & $\tilde{G}$ & (same) \\
\hline
\end{tabular}
\end{indented}
\end{table}

\par From Tab.~\ref{tab:1} we can see that the particle content has doubled from the SM. However, the SUSY does not determine the masses of the sparticles, even though all dimensionless couplings with these particles (such as Yukawa and four point couplings) are. The relationships between couplings can be understood only from the full supersymmetric theory.

\par
\begin{figure}[h]
\centering
\includegraphics[width=9cm]{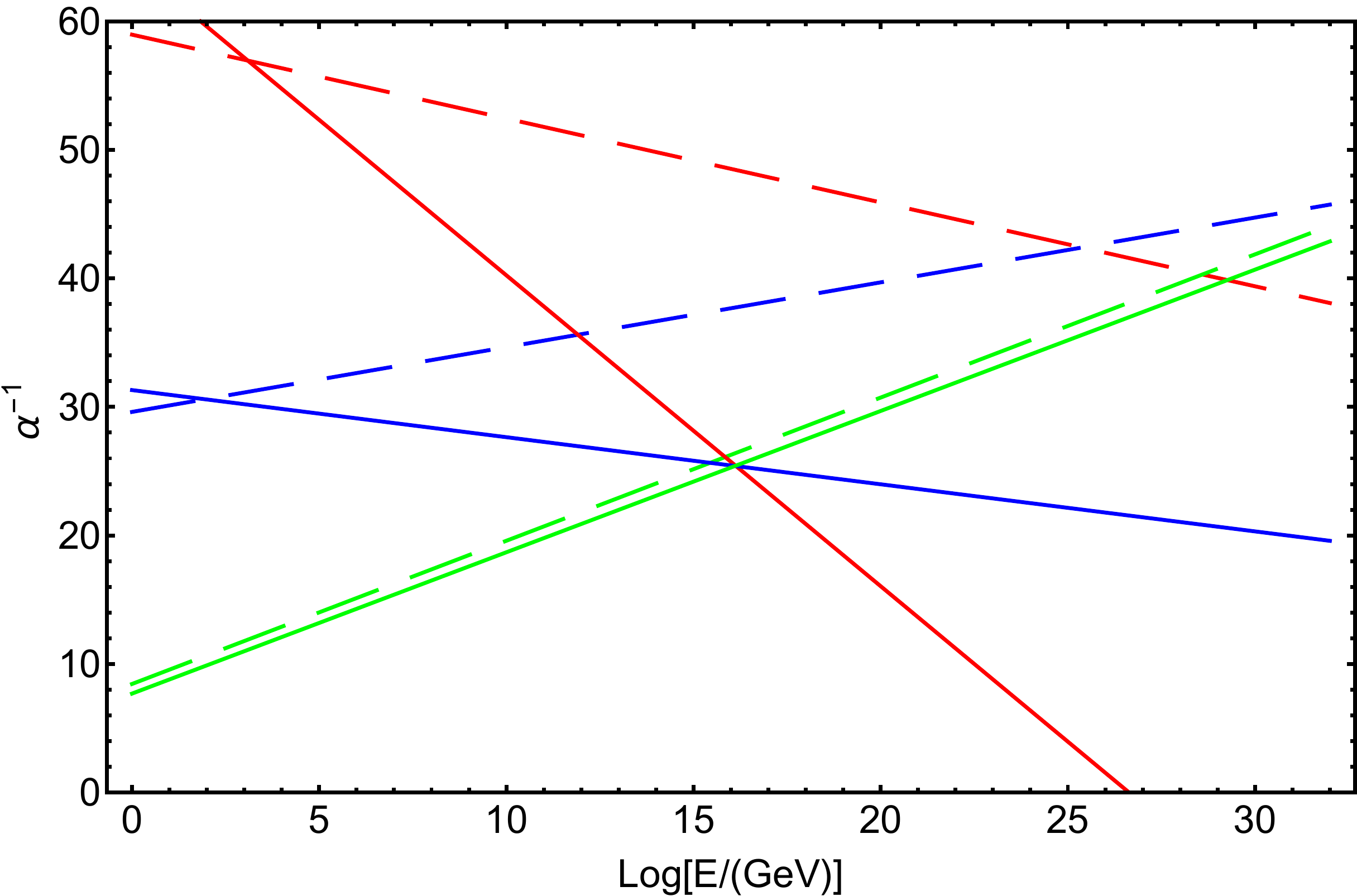}
\caption{\it The one loop renormalisation group evolution of the gauge couplings in the SM (dashed lines) and MSSM (solid lines).}\label{fig:2}
\end{figure}

 Let us consider some features of supersymmetric models:
\begin{itemize}
\item[--] As the quadratic divergence arising from the top loop is cancelled by the stop loop (from a Higgs-Higgs-stop-stop interaction), given that both diagrams are proportional to $y_t^2$, etc., there are no quadratic divergences in this theory. However, as scalar particles are in the same multiplet as fermions (fermion mass being logarithmically divergent), the Higgs quartic coupling is proportional to the square of the higgsino-gaugino loops. Therefore fine-tuning within the Higgs sector is greatly reduced.
\item[--] As the four point coupling of the Higgs is now a gauge coupling, it is always positive at the Planck scale. Please see the work, including proceedings, of Abdalgabar {\it et al.} \cite{Abdalgabar:2014bfa} and references therein.
\item[--] From gauge invariance there are no baryon and lepton number violating processes in the SM. This is not the case in SUSY models as higgsinos carry the same quantum numbers as leptons.
\item[--] Gauge coupling: In models such as the MSSM the number of particles can be doubled from the SM, with runnings of the gauge couplings being modified above the sparticle mass scale. As such the gauge couplings unify much better than in the SM case at the GUT scale (see Fig.~\ref{fig:2}). This means that a supersymmetric GUT agrees with current experimental results, even though fine-tuning issues may persist, as the Higgs sector may violate the GUT symmetry.
\end{itemize}

\newpage
\subsection{Origins of SUSY breaking}

\par As a final discussion point on SUSY, models like the MSSM are incomplete theories as the necessary SUSY breaking mechanism comes from elsewhere. Note the SUSY breaking set-up has, in general, some hidden sector containing fields which spontaneously break the SUSY. The hidden sector does not couple directly to the visible sector but does so through some messenger sector, where the messenger particles have some mass scale.

\par However, this messenger sector is already severely constrained for the MSSM sector from such measurements as flavour changing neutral currents ($K^0 - \bar{K}^0$ mixing for example). Even so, obtaining information on the hidden sector remains difficult. In some cases the gravitino (the graviton's superpartner) mass $m_{3/2} = F_0/M_{Pl}$, where $F_0$ is the total energy of SUSY breaking, could be the lightest supersymmetric partner (LSP). Often the next to LSP (NLSP) would be long-lived in such cases. As such the NLSP may be detectable at colliders, with its life-time giving information on this hidden sector responsible for SUSY breaking. Note that when the gravitino is not the LSP, the gravitino will be long enough lived to effect the big bang nucleosynthesis.

\par The mechanisms of the messenger sector set the scale of the sparticle masses, where the on-shell SUSY particle masses are then determined from the renormalisation group equations being run down to the lower energy scale. A more complete discussion can be found in Ref.~\cite{Hall:1983iz}.

\section{Extra-dimensions}

\par Among the other possible models which address the fine-tuning problem (we shall review) are those with extra-dimensions. These models address this problem by noticing that the observed Planck scale is effective and that the true (higher-dimensional) Planck scale can be smaller and even of the order of the electroweak scale. As such the parameters in the Higgs sector are of the order of the true Planck scale and not the effective one. Furthermore, in some extra-dimensional models the Higgs may be the fifth-dimensional component of the gauge field, and as such its parameters are protected by the gauge symmetry (no quadratic divergences).

\par To given an overview of some of the ideas used in these models, let us consider the case where we have additional spatial dimensions that are compactified with radius $R$ \cite{Antoniadis:1998ig}. If we have one flat extra-dimension, then fields propagating in this extra-dimension must obey a periodic boundary condition, for example
\begin{eqnarray}
\phi (x, y) & = & \phi (x, y+R) \; ,
\end{eqnarray}
with $x$ being our usual four spacetime dimensions, and $y$ our extra-dimension. 

\newpage
\par With such a boundary condition wavefunctions can be written as
\begin{eqnarray}
\psi (x, y) & = & \psi'(x) \mathrm{exp}(ip_5y) \; ,
\end{eqnarray}
where $p_5$ is the fifth component of our 5-momenta and satisfies $p_5 R = 2\pi n$ for an integer $n$. The equation of motion for a particle moving in the additional dimension becomes
\begin{eqnarray}
E_n^2 & = & p^2 + p_5^2 = p^2 + (2\pi)^2 \left( \frac{n}{R} \right)^2 \; .
\end{eqnarray}
That is, we have an infinite tower of massive particles in the effective four-dimensional theory, with masses equal to the discrete values of $p_5^2$.

\par The couplings in the higher-dimensional theory are related to those in the effective four-dimensional case, however, this can be non-trivial, depending on the model. For gauge couplings in our simple case above
\begin{eqnarray}
\int d^4x dx_5 \frac{1}{g_5^2} F_{\mu\nu}F^{\mu\nu} & \to & \int d^4x \frac{1}{g_4^2} F_{\mu\nu}F^{\mu\nu} \; ,
\end{eqnarray}
where $g_4 = g_5/\sqrt{R}$. So as the extra-dimension becomes larger, $g_4$ is reduced. This is also true for gravitational interactions. The four-dimensional gravitational interaction may be very weak if the size of the extra-dimensions are very large. In the case of large extra-dimensional models these can solve the fine-tuning problem as discussed above, by making the true Planck scale in the higher-dimensional theory much smaller.

\par There are many other varieties of extra-dimensional models, and they are not all flat. A famous example is that of the Randall-Sundrum model \cite{Randall:1999ee} where the additional spatial dimension has the ``warped" metric
\begin{eqnarray}
ds^2 & = & e^{-2\sigma(\phi)} \eta_{\mu\nu} dx^\mu dx^\nu + r_c^2 d\phi^2 \; ,
\end{eqnarray}
and where the boundaries of our additional dimension are $\phi = 0$ and $\pi$. In this case the action becomes
\begin{eqnarray}
S_{gravity} & = & \int d^4 x \int^{+\pi}_{-\pi} d\phi \sqrt{-G} - \Lambda + 2 M^3 R \; ,
\end{eqnarray}
with ``warp" factor
\begin{eqnarray}
\sigma(\phi) & = & r_c |\phi| \sqrt{\frac{-\Lambda}{24M^3}} \; , 
\end{eqnarray}
together with appropriate actions on the boundaries.

\section{Concluding remarks}

\par As a final thought, note that there are other indications of the existence of new physics between the weak scale and the Planck scale. Firstly,  consider the Higgs potential and how it evolves with energy in the SM, in particular its stability. The potential is a function of the top and Higgs masses, and current top and Higgs mass measurements favour a metastable Higgs potential \cite{Alekhin:2012py}. Now, there is no reason that the Higgs vev should fall in such a metastable region, and this also suggests that additional particles that couple to the Higgs sector change the shape of the potential dramatically, see Refs.~\cite{Liu:2012mea} for example. So further analysis of this region of parameter space may give indications of BSM physics.

\par Though we have not discussed it in this review, we note that another topical result which requires some new physics is dark matter. From various cosmological and astrophysical observations we know that $\sim 27$\% of the energy content of the Universe is in the form of dark matter. And therefore such particles must be stable and neutral. Many candidates for dark matter particles exist in a range of BSM theories, including the MSSM, but more studies are required and this is an ongoing field of research.

\par The points that have been raised in this brief proceedings will hopefully motivate the participants of this workshop to further readings on these and related topics, such as Refs.~\cite{Hall:1983iz, Rattazzi:2003ea, Arneodo:2013re, Strege:2012bt}.


\end{document}